\documentstyle[12pt]{article}

\def\be{\begin{equation}}
\def\ee{\end{equation}}
\def\a{\alpha}

\def\G{\Gamma} 
\def\sin{\mbox{sin}}
\def\exp{\mbox{exp}}
\def\ra{\rangle}
\def\la{\langle}
\def\Im{\mbox{Im}}
\def\sign{\mbox{sign}} 
\def\low{\mbox{low}}
\def\w{\omega}
\def\d{\delta}
\def\e{\epsilon}
\def\ha{\hat{a}} 
\def\hc{\hat{c}} 
\def\pp{p^{\prime}} 
\def\pq{q^{\prime}}

\begin{document}

\begin{center}
{\bf Threshold singularities in the correlators of the one-dimensional models.} 
\end{center}
\vspace{0.2in}
\begin{center}
{\large A.A.Ovchinnikov}
\end{center}

\begin{center}
{\it Institute for Nuclear Research, RAS, Moscow}
\end{center}

\vspace{0.1in}

\begin{abstract}

 We calculate the threshold singularities in one-dimensional 
models using the universal low-energy formfactors obtained 
in the framework of the non-linear Luttinger liquid model. 
 We find the reason why the simplified picture of the impurity 
moving in the Luttinger liquid leads to the correct results. 
 We obtain the prefactors of the singularities including their 
$k$- dependence at small $k<<p_F$. 

\end{abstract}

\vspace{0.2in}

{\bf 1. Introduction}

\vspace{0.2in}

Although the results for the large distance behaviour of correlators 
in one-dimensional quantum systems are known for a long time \cite{LP}, 
\cite{H}, the calculation of 
the dynamical correlation functions and in particular the calculation 
of the exponents and the amplitudes of the threshold singularities in the 
$(\w,k)$ plane for different correlators was performed only recently 
\cite{S},\cite{IG},\cite{Shashi}. In ref.\cite{S} the limit of small 
momentum is considered, while in ref.\cite{IG},\cite{Shashi} the critical 
exponents at arbitrary momentum are calculated. 
The results are different from the predictions of the usual 
Luttinger liquid theory and were obtained in the framework of the so called 
non-linear Luttinger liquid theory. 
In the present paper we show that although in \cite{S}   
the final results for the critical exponents are correct, their derivation 
in the framework of the non-linear Luttinger liquid theory is not rigorous 
and present the rigorous way to calculate the form of the threshold 
singularities using the sum over the formfactor series. 
We show why the simplified treatment of ref.\cite{S} 
leads to the correct results for the critical exponents.  
While the universal method of ref.\cite{S} gives the correct values of the 
critical exponents, it fails to predict the prefactors of the 
singularities. A different method of refs.\cite{IG},\cite{Shashi} gives the 
correct predictions for the prefactors. 
However in \cite{IG},\cite{Shashi} 
the notion of the mobile impurity coupled to the Luttinger liquid is 
only the hypothesis which should be confirmed by the rigorous derivation. 
Although the formfactor method was recently used in ref.\cite{K} to obtain 
the correlators in the Lieb-Liniger Bose gas we find it useful to present 
the derivation starting from the simple particle-hole low-energy formfactors 
obtained as a matrix elements of the exponential operators in the framework 
of the Luttinger liquid model (the authors of ref.\cite{K} used the formfactors 
obtained in the other papers \cite{K1} which have a very 
complicated form). Here the drastic simplification is achieved due to the 
assumption $k<<p_F$, where $p_F$ is the Fermi momentum, which leads to the 
simple and transparent derivation of various exponents in the framework 
of an arbitrary gapless one-dimensional model or the non-linear Luttinger 
liquid. We believe that the simple derivation of the formfactors for the 
high-energy particle from the universal low -energy expression for the 
formfactors, connected with the exponential operator, is the useful result 
which allows one to find the corresponding exponents in a transparent way 
and make the connection with the simplified approach of ref.\cite{S}. 
Let us stress that the formfactors obtained in ref.\cite{K1} 
for the exactly-solvable model do not rely on the universal low-energy 
expression for the formfactors and have the complicated form. 
We also propose the simple way to calculate the sum over the low-energy 
formfactors. In addition we find the $k$- dependence of the prefactors of 
the singularity at $k<<p_F$ and connect the prefactor with the prefactor 
of the large distance asymptotics of the corresponding correlator.

\vspace{0.2in}

{\bf 2. Formfactors in the non-linear Luttinger liquid.}

\vspace{0.2in}

To compare the method of calculating the critical exponents
employed in ref.\cite{S} with the method yielding the correct predictions 
for the prefactors of the threshold singularities 
in the non-linear Luttinger liquid,  
let us briefly review the method of ref.\cite{S}. 
Let $a^{+}_{k}$ ($a_{k}$), $c^{+}_{k}$ ($c_{k}$) to be the operators of 
particles at the right and the left branches of the Luttinger model. 
The Hamiltonian of the non-linear Luttinger liquid has the following form in 
terms of the operators of quasiparticles 
(for example, see \cite{ML},\cite{R},\cite{RMP}): 
\[
H_{NL}=\sum_{k}vk(\ha^{+}_{k}\ha_{k}-\hc^{+}_{k}\hc_{k})+
\sum_{k}(k^2/2m)(\ha^{+}_{k}\ha_{k}+\hc^{+}_{k}\hc_{k}),  
\]
where $\ha$ ($\hc$) - are the operators of quasiparticles at the right (left) 
branch and $v$ and $m$ are the parameters of the model. 
The quasiparticle operators are connected with the particle operators 
as follows: 
\be
a^{+}(x)=\ha^{+}(x)F^{+}(x), ~~~~~~
F^{+}(x)=\exp\left(i2\pi\d_{1}\hat{N}_{1}(x)+i2\pi\d_{2}\hat{N}_{2}(x)\right), 
\label{F}  
\ee
where the operators $\hat{N}_{1,2}(x)$ are related to the usual Luttinger liquid 
operators 
\[
N_{1,2}(x)=\frac{i}{L}\sum_{p\neq 0}\frac{\rho_{1,2}(p)}{p}e^{-ipx}, ~~~~
\rho_{1}(p)=\sum_{k}a^{+}_{k+p}a_k, ~~~~\rho_{2}(p)=\sum_{k}c^{+}_{k+p}c_k, 
\]
via the Bogoliubov rotation 
$N_{1}(x)+N_{2}(x)=(1/\sqrt{\xi})(\hat{N}_{1}(x)+\hat{N}_{2}(x))$. 
Here $\xi$ is the usual Luttinger liquid parameter. The most simple way to 
obtain the equation (\ref{F}) is to use Bosonization. The parameters $\d_{1,2}$ 
in (\ref{F}) are given by the equations: 
\be
\d_{1}=1-\frac{1}{2}\left(\sqrt{\xi}+1/\sqrt{\xi}\right),~~~~~
\d_{2}=\frac{1}{2}\left(\sqrt{\xi}-1/\sqrt{\xi}\right). 
\label{delta}
\ee
Clearly, the intermediate states in the correlator depending on the 
energy $\w$ and the momentum $k$ which are responsible for the threshold 
singularities are the eigenstates with the single high-energy quasiparticle 
with the momentum $p\sim k$ and the low-energy particle-hole excitations 
near the left and the right Fermi- points with the momenta $q_{1,2}<<k$. 
Then the main assumption of ref.\cite{S} is the factorization 
of the corresponding matrix element of the following form: 
\be
\la p, \low(q_1,q_2)|a^{+}|0\ra=\la p, \low(q_1,q_2)|\ha^{+}F^{+}|0\ra=
\la p|\ha^{+}|0\ra \la \low(q_1,q_2)|F^{+}|0\ra. 
\label{factor} 
\ee
Using this equation one can easily obtain in the 
spectral density the threshold singularities of the form $(\w-\e(k))^{-\mu}$ 
with $\mu=1-\d_1^2-\d_2^2$ for the particle excitation 
(particle at the momentum $k>0$) and $ \mu=1-(2-\d_1)^2-\d_2^2 $ for the hole 
excitation (hole at the momentum $k<0$). 
Although these results for the critical exponents 
are correct (see below) one can see that the basic 
assumption for the formfactors (\ref{factor}) is obviously not correct. 
In fact, in the framework of bosonization $a^{+}$ is the exponential operator 
(this is valid also for the non-linear Luttinger liquid) 
and for $p<<p_F$ the average of the type (\ref{factor}) with an arbitrary number 
of the particle-hole excitations are well known \cite{Shashi},\cite{K1},\cite{O}.  
In fact the result for the exponential operator has the following form \cite{A}:  
\be
\la p_i,q_i|e^{a\frac{2\pi}{L}\sum_{p>0}\frac{\rho_1(p)}{p}}|0\ra=F_a(p_i,q_i)= 
\frac{\prod_{i<j}(p_i-p_j)\prod_{i>j}(q_i-q_j)}{\prod_{i,j}(p_i-q_j)}
\prod_if^{+}(p_i)\prod_if^{-}(q_i), 
\label{ff}
\ee
\[
f^{+}(p)=\frac{\G(p+a)}{\G(p)\G(a)},~~~~~
f^{-}(q)=\frac{\G(1-q-a)}{\G(1-q)\G(1-a)}, 
\] 
where integers $p_i>0$ are the positions of particles, $q_i\leq 0$ are the 
positions of the holes and the parameter $a=(1/2)(\sqrt{\xi}+1/\sqrt{\xi})$ 
for the contribution of the right-moving particles in the matrix element 
(\ref{factor}) (we consider the system of Fermi- particles in the vicinity of 
the momentum $+p_F$ in Sections 2,3). 
Below we show how the expression for the formfactors (\ref{ff}) leads to the 
correct expressions for the threshold singularities in the non-linear Luttinger 
liquid and explain why the naive approximation (\ref{factor}) leads to the 
correct results for the critical exponents.

\vspace{0.2in}

{\bf 3. Formfactors for the eigenstates with high energy particle.} 

\vspace{0.2in}

In this section we find the general expression for spectral density in the 
framework of the non-linear Luttinger liquid theory. 
Let us begin with the particle excitation. 
Let us find the formfactor of the operator $a^{+}(0)$ (\ref{factor}) for 
the case of the single high-energy particle with the momentum $p>>q_{1,2}$ at 
the right branch. Let us denote by $p_i$ and $q_i$ the positions of the 
particles and the holes at low energy (in the state $|\low(q_1)\ra$  
in eq.(\ref{factor})). Then the part of the formfactor (\ref{factor}) 
corresponding to the right branch takes the following form: 
\be
\la p,\{p_i,q_i\}|O_{1}^{+}K_{1}^{+}|0\ra=\la p,\{p_i,q_i\}|O_{1}^{+}|1\ra=
\la\{\pp,\pp_i\},\{\pq_i,0\}|O_{1}^{+}|0\ra, 
\label{me}
\ee
where the operator $O_{1}^{+}=\exp(-i\pi(\xi^{1/2}+\xi^{-1/2})\hat{N}_{1})$,  
$K_{1}^{+}$ is the Klein factor, $|1\ra$ is the state with the single extra 
particle at the right Fermi- point and $\{\pp,\pp_i\}$ ($\{\pq_i,0\}$)- 
are the positions of particles (holes) in the final matrix element of the 
operator $O_{1}^{+}$. Clearly we have: 
\[
\pp=p-1,~~~~\pp_i=p_i-1,~~~~\pq_i=q_i-1. 
\]
Substituting these momenta into the basic relation for the formfactors 
(\ref{ff}) and taking into account that $p>>p_i,q_i$, we obtain at 
$a=(1/2)(\xi^{1/2}+\xi^{-1/2})$ the simple expression for the matrix 
element (\ref{me}). In fact, lat us consider the Cauchy determinant in 
eq.(\ref{ff}) and use the following simple equations: 
\[
\frac{1}{p_i-1}\frac{\G(p_i-1+a)}{\G(p_i-1)}=\frac{\G(p_i+(a-1))}{\G(p_i)}, 
~~
(1-q_i)\frac{\G(1-(q_i-1)-a)}{\G(1-(q_i-1))}=\frac{\G(1-q_i-(a-1))}{\G(1-q_i)}. 
\]
Then apart from the simple factor $f(p)$ we obtain the function 
$F_{a^{\prime}}(p_i,q_i)$ (\ref{ff}) with the value of the exponent 
$a^{\prime}=a-1$ which corresponds exactly to the matrix element of 
the operator $F^{+}$ (\ref{F}) over the eigenstate $\la p_i,q_i|$ 
with the exponent $\d_1$ given by the equation (\ref{delta}). 
Clearly the part of the formfactor 
corresponding to the left branch is given by the matrix element 
of the operator (\ref{F}) with the exponent $\d_2$. Using this result 
one can easily obtain the threshold singularity for the particle 
excitation with the exponent $\mu=1-\d_1^2-\d_2^2$.   
Before performing this calculation let us comment on the singularity 
corresponding to the hole excitation (at $\w>0$). 
In his case the eigenstate which is responsible for the singularity 
is given by the state with a single hole located at the momentum $-k$ 
(which corresponds to the contribution to the total momentum $k$) and 
the two particles located at the right Fermi- point. 
The calculation of the formfactors is analogous to the previous 
case and leads to the exponents $\tilde{\d}_1=2-\d_1$, $\tilde{\d}_2=\d_2$ 
in the analog of the $F$ - operator (see (\ref{factor})). Clearly this leads 
to the exponent $\mu=1-(2-\d_1)^2-\d_2^2$ for the threshold 
singularity for the hole excitation. 
To summarize we write down the explicit expressions for the 
formfactors for the eigenstate with the high-energy particle (hole): 
\[
\la p,\{p_i,q_i\}|O_{1}^{+}|1\ra=f(p)F_{a-1}(p_i,q_i)\la0|O_1|0\ra, 
\]
\[
\la p_0,\{p_i,q_i\},2|O_{1}^{+}|1\ra=
\tilde{f}(p_0)F_{a+1}(p_i-2,q_i-2)\la0|O_1|0\ra, 
\]
where the smooth functions $f(p)$, $\tilde{f}(p)$ are defined below. 
Here the parameter $\d_1$ is related to the parameter $a$ as $\d_1=1-a$. 
Note that since it the latter case we have two additinal particles 
at the right Fermi- point the sum over the quantum numbers 
$p_i$, $q_i$ goes for $p_i>2$, $q_i\leq 2$ such that as a result 
we obtain the standard sum over the low-energy states. 
One can see that the function $F_{a-1}(p_i,q_i)$ corresponds exactly 
to the operator $F^{+}$ (\ref{F}) which is the reason why the 
the simplified approach of ref.\cite{S} leads to the correct results. 
In other words since the formfactor is given by the matrix element 
of the exponential operator one finds that the extraction of the 
quasiparticle (\ref{F}) is equivalent to the substitution 
$a\rightarrow a-1$ in the formfactor. It is an interesting fact 
that if we consider the dependence on the momentum of the high-energy 
particle we get the formfactor corresponding to the different 
operator. Note that since $f(p)$ is the 
smooth function one can neglect its dependence on $q_{1,2}$ and 
perform the summation over the low-energy particle-hole states 
independently taking $f^{2}(k)$ as the common factor. 
Let us show how the results for the formfactors with the high- 
energy particle (hole) lead to the expressions for the threshold 
singularities in the non-linear Luttinger liquid. 
Here we consider the Fourier transform of the correlator 
$\la a(x,t)a^{+}(0)\ra$. Bosonization allows to express the operator 
$a(x)$ through the standard exponential operators $O(x)$ so that we 
should calculate the correlator  
$\int dxdte^{i\w t-ikx}\la O(x,t)O^{+}(0)\ra$. For example, for the 
particle case, inserting the complete set of states with the single 
high-energy particle with the momentum $p$ we obtain the following 
expression for this correlator: 
\[
\sum_{p,\low(q_1,q_2)}\d_{k,p+q_1-q_2}2\pi L\d(\w-\epsilon_1(p)-v(q_1+q_2)) 
|\la p,\low(q_1,q_2)|O^{+}K_1^{+}|0\ra|^2,  
\]
where the formfactor was calculated above. 
First, consider the case of the particle 
excitation. Substituting the formfactors found above into the sum over 
the formfactors and performing the sum over the position of the 
high-energy particle $p=k-q_1+q_2$, we obtain the expression: 
\be
2\pi L\sum_{q_1,q_2>0}\d(\d\w+C_{1}q_1-C_{2}q_2)f^2(k)
F(q_1,\d_1^2)F(q_2,\d_2^2)|\la0|O|0\ra|^2, 
\label{NLp}
\ee
where we denote $C_1=v_d-v=k/m$ ($v_d=v+k/m$),  
$C_2=v_d+v=2v+k/m$ and $\d\w=\w-\e_1(k)$, $\e_1(k)=vk+k^2/2m$. 
The functions 
$f(k)$, $F(q_1,\d_1^2)$, $F(q_2,\d_2^2)$ which enter the equation 
(\ref{NLp}) are given by the following equations. 
First, the function $f(k)$ is given by 
\[
f(k)=\frac{1}{\bar{k}}\frac{\G(\bar{k}+a)}{\G(\bar{k})\G(a)}\simeq 
\frac{1}{\G(a)}\bar{k}^{a-1},  ~~~~ \bar{k}=\frac{Lk}{2\pi}.
\]
The function $\tilde{f}(k)$ is given by the same expression with the 
replacement $a\rightarrow-a$. 
Second, the functions $F(q,a^2)$ which are the sums over the low-energy 
particle-hole states of the corresponding operators, 
are given by the following equation: 
\[
F(m,a^2)=\sum_{n}\sum_{p_i,q_i,p-q=m}|F_a(p_i,q_i)|^2= 
\frac{\G(a^2+m)}{\G(m+1)\G(a^2)}, 
\]
where $p=\sum_{i=1}^{n}p_i$, $q=\sum_{i=1}^{n}q_i$ 
and the function $F_a(p_i,q_i)$ is defined in eq.(\ref{ff}).
For the hole excitation after performing the sum over the position of 
the hole we obtain the expression similar to the equation (\ref{NLp}): 
\be
2\pi L\sum_{q_1,q_2>0}\d(\d\w-C_{1}q_1-C_{2}q_2)\tilde{f}^{2}(k)
F(q_1,\tilde{\d}_1^2)F(q_2,\tilde{\d}_2^2)|\la0|O|0\ra|^2, 
\label{NLh}
\ee
where $\tilde{\d}_1=2-\d_1$, $\tilde{\d}_2=\d_2$ and 
$\d\w=\w-\e_2(k)$, $\e_2(k)=vk-k^2/2m$.  
Below we write down the expressions for the spectral density for 
an arbitrary gapless model of Fermi- particles and calculate the 
expressions (\ref{NLp}), (\ref{NLh}).

\vspace{0.2in}

{\bf 4. Threshold singularities for an arbitrary model of Fermi- particles.}  

\vspace{0.2in}

In this section we calculate the threshold singularities in the spectral 
density for an arbitrary model of Fermi- particles. 
Here we re-derive in a simple way the results obtained earlier in 
ref.\cite{Shashi} and later confirmed in ref.\cite{K}. 
We also show that the 
prefactors of the singularities are related to the prefactors of the leading 
large- distance asymptotics of the correlators of the corresponding 
operators. The spectral density of the model is defined as 
$A(\w,k)=(1/\pi)\sign(\w)\Im G(\w,k)$, where $G(\w,k)$ - 
is the standard Green function of the model. In terms of the Fermi- fields 
$\psi(x,t)$ ($\psi^{+}(x,t)$) it has the following form: 
\be 
A(\w,k)=\frac{\sign(\w)}{2\pi}\int dxdt
\left(e^{i\w t-ikx}\la\psi(x,t)\psi^{+}(0)\ra+ 
e^{-i\w t+ikx}\la\psi^{+}(x,t)\psi(0)\ra\right). 
\label{A}
\ee
The case $\w>0$ corresponds to the first term in eq.(\ref{A}) and 
the case $\w<0$ corresponds to the second term in eq.(\ref{A}). 
Since in the framework of the effective low-energy 
theory which is given by the non-linear Luttinger liquid we have the 
equality $\psi^{+}(x)=CO^{+}(x)K_1^{+}$, where $C$ is some constant 
(the operator $a^{+}$ is related to the exponential operator $O^{+}$ 
in standard way) 
the expressions for $A(\w,k)$ can be easily obtained from the equations 
(\ref{NLp}), (\ref{NLh}). One should take into account the normalization  
in eq.(\ref{A}) and substitute the lowest formfactor $\la1|\psi^{+}(0)|0\ra$,  
where we denote by $\la1|$ the eigenstate with the one extra particle at the 
right Fermi- point, for the average $\la O\ra$. In this way we find 
in the thermodynamic limit the following expressions for the spectral 
density. For the particle excitation  
\be 
A(\w,k)=L^{3}f^2(k)\frac{1}{(2\pi)^2}\frac{1}{\G(\d_1^2)\G(\d_2^2)} 
\int_{0}^{\infty}dq_1 dq_2\d(\d\w+C_1q_1-C_2q_2)
\frac{1}{\bar{q}_1^{1-\d_1^2}}\frac{1}{\bar{q}_2^{1-\d_2^2}} 
|\la1|\psi^{+}|0\ra|^2, 
\label{Ap}
\ee
where $\bar{q}_{1,2}=Lq_{1,2}/2\pi$ and for the hole excitation 
\be 
A(\w,k)=L^{3}\tilde{f}^{2}(k)\frac{1}{(2\pi)^2}
\frac{1}{\G(\tilde{\d}_1^2)\G(\tilde{\d}_2^2)} 
\int_{0}^{\infty}dq_1 dq_2\d(\d\w-C_1q_1-C_2q_2)
\frac{1}{\bar{q}_1^{1-\tilde{\d}_1^2}}
\frac{1}{\bar{q}_2^{1-\tilde{\d}_2^2}} 
|\la1|\psi^{+}|0\ra|^2. 
\label{Ah}
\ee
Combining the degrees of the length $L$ in these equations we find that 
the dependence on $L$ is contained in the factor 
$L^{\a}|\la1|\psi^{+}|0\ra|^2$, where the exponent 
\[
\a=-1+2a+\d_1^2+\d_2^2=\frac{1}{2}\left(\xi+\frac{1}{\xi}\right)  
\]
is proportional to the scaling dimension of the operator $\psi$. 
That means that the factor $L^{\a}|\la1|\psi^{+}|0\ra|^2$ is independent 
of $L$ (for example, see \cite{Shashi},\cite{O}) which gives the important 
consistency check of the whole method. The similar relation with the 
replacement $a\rightarrow-a$ holds for the hole excitation. 
The integrals in the equations (\ref{Ap}), (\ref{Ah}) can be easily 
calculated which leads to the following expressions for the threshold 
singularities at $\w>0$. For the particle excitation we have: 
\be 
A(\w,k)=\frac{(2\pi)^{1-\a}}{\pi}\frac{1}{k^{2-2a}}\frac{1}{\G^2(a)} 
\frac{\G(1-\d_1^2-\d_2^2)(L^{\a}|\la1|\psi^{+}|0\ra|^2)}{(v_d-v)^{\d_1^2} 
(v_d+v)^{\d_2^2}}
\label{part}
\ee
\[
~~~~~~~~~\times \left(\theta(\d\w)\sin(\pi\d_2^2)+
\theta(-\d\w)\sin(\pi\d_1^2)\right)\frac{1}{|\d\w|^{1-\d_1^2-\d_2^2}},  
\]
where $a=(\xi^{1/2}+\xi^{-1/2})/2$. Here the spectral density is not equal 
to zero at both sides of the point $\e_1(k)=vk+k^2/2m$. 
For the hole excitation we have: 
\be
A(\w,k)=(2\pi)^{1-\a}\frac{1}{k^{2+2a}}\frac{1}{\G^2(-a)} 
\frac{(L^{\a}|\la1|\psi^{+}|0\ra|^2)}{\G(\tilde{\d}_1^2+\tilde{\d}_2^2)
(v-v_d)^{\tilde{\d}_1^2}(v_d+v)^{\tilde{\d}_2^2}}\frac{\theta(\d\w)}
{(\d\w)^{1-\tilde{\d}_1^2-\tilde{\d}_2^2}}. 
\label{hole}
\ee
Here the spectral density is not equal to zero only above the point 
$\e_2(k)=vk-k^2/2m$. In both cases we have singled out the combination 
$L^{\a}|\la1|\psi^{+}|0\ra|^2$ which is independent of $L$ 
in thermodynamic limit. In fact one have the following relation 
for the prefactor $C_0$ of the leading asymptotic behaviour of the 
Green function: 
\[
2(L/2)^{\a}|\la1|\psi^{+}|0\ra|^2=C_0.  
\]
Here the exact definition of the constant $C_0$ is given by the equation: 
$\la\psi^{+}(x)\psi(0)\ra\simeq\sin(p_{F}x)C_{0}/(\pi x)^{\a}$. 
Thus both functions (\ref{part}), (\ref{hole}) have the correct 
thermodynamic limit. One can express the prefactors of the singularities 
in the equations (\ref{part}), (\ref{hole}) through the formfactor 
$\la k|\psi^{+}|0\ra$, where the eigenstate $\la k|$ is the state with 
the single extra qusiparticle with the momentum $k$. In fact using the 
universal particle-hole formfactors one can easily obtain:  
\[
|\la k|\psi^{+}|0\ra|^2=
\frac{1}{k^{2-2a}}\left(\frac{2\pi}{L}\right)^{2-2a}\frac{1}{\G^2(a)}
|\la1|\psi^{+}|0\ra|^2.  
\]
This equation can be substituted into the equations 
(\ref{part}), (\ref{hole}) which absorbs the explicit dependence of the 
prefactors on the momentum $k<<p_F$ which is now contained in the 
formfactor $\la k|\psi^{+}|0\ra$. Let us note that if the prefactors  
are expressed through the formfactor $\la k|\psi^{+}|0\ra$, then 
the results (\ref{part}), (\ref{hole}) are valid for an arbitrary 
$k\sim p_F$. The results (\ref{part}), (\ref{hole}) 
agree with the results obtained in ref.\cite{Shashi}.   
The results presented above refer to the case $\w>0$ ($k>0$). 
For the 
case $\w<0$ one should consider the matrix element $\la n|a|0\ra$, 
where $\la n|$ is the excited state. 
For the initial model of the Fermi- particles the momentum of 
this state is $-p_F-k$ ($k>0$). For the particle excitation 
the eigenstate $\la n|$ is now given by the state with the 
particle with the momentum $-k$ at the left branch and the two 
holes at the left and the right Fermi- points. 
For the hole excitation one should consider the state with the 
hole with the momentum $-k$ at the left branch, particle at the 
left Fermi- point and the hole at the right Fermi- point. 
The formfactors corresponding to the left branch are given 
by the expression similar to eq.(\ref{ff}).  
Repeating the calculations presented in Section 3 one easily 
finds that the exponents of the threshold singularities 
are given by the following values of the parameters: 
\[
a=\frac{1}{2}\left(\frac{1}{\sqrt{\xi}}-\sqrt{\xi}\right), ~~~
\d_1=\tilde{\d}_1=
\frac{1}{2}\left(\sqrt{\xi}+\frac{1}{\sqrt{\xi}}\right), ~~~
\d_2=2-\tilde{\d}_2=1-a. 
\]
In this case one can easily obtain the expressions similar to 
(\ref{part}), (\ref{hole}).

Let us calculate the threshold singularities for the 
density structure factor (density-density correlator). 
The first term in the bosonization expansion of the density 
operator is $\rho(x)=(a^{+}a+c^{+}c)(x)$. 
At low momentum $k<<p_F$ the next terms of the type 
$(a^{+}c)$, $(c^{+}a)$ corresponding to the high momenta 
$\pm 2p_F$ can be omitted. After the Bogoliubov rotation 
we obtain the density in terms of the quasiparticle operators: 
$\rho(x)=(1/\sqrt{\xi})(\ha^{+}\ha+\hc^{+}\hc)(x)$. 
From this expression one can see that the density structure 
factor is given by the step function 
$(m/k\xi)\theta(\w-\epsilon_2(k))\theta(\epsilon_1(k)-\w)$, 
where $\epsilon_{1,2}(k)=vk\pm k^2/2m$. 
This function does not have the power-law singularities which 
means that in the limit of small $k<<p_F$ the exponents $\mu$ 
are equal to zero. This result is in agreement with the results 
of ref.\cite{Shashi},\cite{K}, 
The behaviour of the density structure factor 
at $k\sim p_F$ is beyond the scope of the present Letter.

As an example, let us calculate the threshold singularities at the 
momentum $2p_F+k$, where $k<<p_F$. To calculate this correlator one needs 
the matrix elements of the form $\la n|(a^{+}c)|0\ra$, 
where $\la n|$ is the excited state. 
Here one should consider the eigenstate with the particle with the  
momentum $k$ at the right branch and the hole at the left Fermi- point 
for the particle excitation and the eigenstate with the hole with the 
momentum $k$, two particles at the right Fermi- point and the hole 
at the left Fermi- point for the hole excitation. 
One can see that the results for the particle and the hole excitations  
can be easily obtained from the equations (\ref{part}), (\ref{hole}) by 
the substitution of the correct parameters $a$, $\d_{1,2}$.  
The calculation of the formfactors in both cases is analogous to the 
case of the spectral density presented above. Using Bosonization 
one can easily see that now the values of the parameters are: 
\[
\d_1=1-1/\sqrt{\xi}, ~~~\d_2=1/\sqrt{\xi}, ~~~a=1/\sqrt{\xi}, ~~~
\a=2/\xi. 
\]
We find that the condition of the cancellation of powers of $L$,  
$\a=-1+2a+\d_1^2+\d_2^2=2/\xi$ is fulfilled. 
Thus up to a simple factor $2\pi$ the results for the density structure 
factor coincide with the expressions (\ref{part}), (\ref{hole}) with 
the parameters $a$, $\a$, $\tilde{\d}_{1}=2-\d_{1}$, 
$\tilde{\d}_{2}=\d_{2}$ presented above 
with the obvious substitution of the formfactor $\la1|\psi^{+}|0\ra$ 
by the formfactor $\la1,-1|\rho(0)|0\ra$, where $\rho(x)$ is the density 
operator and the state $\la1,-1|$ is the eigenstate with the single 
extra particle at the right Fermi- point and the single hole at the 
left Fermi- point.

Finally, let us consider the system of Bose- particles. It is well known 
that in this case in the framework of Bosonization in the leading order 
the Bose field equals 
$\phi(x)\sim\exp(i\pi\sqrt{\xi}(\hat{N}_1(x)-\hat{N}_2(x)))$.
At $\w>0$ the correlator is determined by the matrix elements of the 
type $\la n|\phi^{+}|0\ra$. 
One can see that in this case all the calculations are exactly 
the same as in the case of the Fermi- particles. Thus we obtain the same 
final expressions (\ref{part}), (\ref{hole}) with the different values 
of the parameters $\d_1$, $\d_2$, $a$, $\a$ given by the equations: 
\[
\d_1=1-\sqrt{\xi}/2, ~~~\d_2=\sqrt{\xi}/2, ~~~a=\sqrt{\xi}/2, ~~~
\a=\xi/2. 
\]
Clearly for the hole excitation the values of the parameters are 
$\tilde{\d}_1=2-\d_1$, $\tilde{\d}_2=\d_2$ as in the case of the 
Fermi- system. Let us note that now the formfactor $\la1|\psi^{+}|0\ra$ 
should by substituted by the formfactor $\la1|\phi^{+}|0\ra$, where 
$\la1|$ is the ground state of the model in the sector with one 
additional particle. 
At $\w<0$ the spectral density is determined by the matrix elements 
$\la n|\phi|0\ra$. The eigenstates are exactly the same as for the case 
of the Fermi- particles. One can easily see that the exponent $\mu$ 
corresponding to the particle (hole) excitation coincides with the 
exponent for the hole (particle) excitation for the case $\w>0$. 
Our results for the system of Bose- particles are in agreement 
with the results of ref.\cite{K},\cite{RMP}.

\vspace{0.2in}

{\bf 5. Conclusion.}  

\vspace{0.2in}

In the present Letter we performed the simple derivation of the 
formfactors for the high-energy particle from the universal low-energy 
expression for the formfactors, connected with the exponential operators 
in the framework of the non-linear Luttinger liquid. 
We performed the summation over the eigenstates responsible for the 
threshold singularities and presented the correct expressions for the 
threshold singularities in the non-linear Luttinger liquid. 
We believe that the simple derivation of the formfactors for the 
non-linear Luttinger liquid as a matrix elements of the exponential 
operators is an interesting result which makes the connection of the 
rigorous approach with the approach of ref.\cite{S}. 
We explained why the simplified approach based on the notion of the 
impurity moving in the Luttinger liquid leads to the correct results 
for the exponents of the threshold singularities. 
Using the formfactor approach we performed the calculation of the 
threshold singularities for the spectral density and the density 
structure factor both for the Bose- and the Fermi- systems.  
Note that once the formfactors for the particle 
and the hole for the operator $a^{+}$ are known the calculations 
for all the other cases become a simple exercise. 
Note that although the results for an arbitrary Bose- system 
could be extracted with some additional assumptions from 
the results of ref.\cite{K}, the results for the Fermi- system 
can be derived rigorously only in the framework of our approach. 
Our results are applicable for an arbitrary gapless one- dimensional 
model. We also found the $k$- dependence of the prefactors 
of the singularities and expressed them through the the prefactors 
of the leading asymptotic behaviour of the correlation functions.

\end{document}